\documentclass[letterpaper, 10 pt, conference]{ieeeconf}  

\IEEEoverridecommandlockouts                              
\usepackage{graphics} 
\usepackage{amsmath} 
\usepackage{amssymb}  
\usepackage{epsfig} 
\usepackage{mathptmx} 
\usepackage{times} 
\usepackage[utf8]{inputenc}
\usepackage{graphicx}
\usepackage{rotating}
\usepackage{caption}
\usepackage{subcaption}
\usepackage{adjustbox}
\usepackage{color} 
\usepackage{xcolor} 
\usepackage{transparent} 
\usepackage{float}
\usepackage{comment}
\usepackage{arydshln}
\usepackage{mathtools}
\usepackage{breqn}
\usepackage{algorithmic}
\usepackage{algorithm}
\usepackage{romannum}
\usepackage{verbatim}
\usepackage[english]{babel}
\newtheorem{definition}{Definition}
\newtheorem{theorem}{Theorem}

\newtheorem{assumption}{Assumption}
\newtheorem{lemma}{Lemma}

\title{\LARGE \bf
Probabilistically safe controllers based on control barrier functions and scenario model predictive control
}

\author{Allan Andre do Nascimento, Antonis Papachristodoulou, Kostas Margellos
\thanks{AAdN, AP and KM acknowledge funding support by MathWorks. AP was supported in part by UK's Engineering, Physical Sciences Research Council projects EP/X017982/1 and EP/Y014073/1.}
\thanks{For the purpose of Open Access, the authors have applied a CC BY public copyright licence to any Author Accepted Manuscript (AAM) version arising from this submission.}
\thanks{All authors are with the Department of Engineering Science,
        University of Oxford, Parks Road, Oxford OX13PJ, United Kingdom.
        {\tt\small \{allan.adn, antonis, kostas.margellos\}@eng.ox.ac.uk}}%
}

\begin{document}

\maketitle
\thispagestyle{empty}
\pagestyle{empty}

\begin{abstract}
Control barrier functions (CBFs) offer an efficient framework for designing real-time safe controllers. However, CBF-based controllers can be short-sighted, resulting in poor performance, a behaviour which is aggravated in uncertain conditions. This motivated research on safety filters based on model predictive control (MPC) and its stochastic variant. MPC deals with safety constraints in a direct manner, however, its computational demands grow with the prediction horizon length. We propose a safety formulation that solves a finite horizon optimization problem at each time instance like MPC, but rather than explicitly imposing constraints along the prediction horizon, we enforce probabilistic safety constraints by means of CBFs only at the first step of the horizon. The probabilistic CBF constraints are transformed in a finite number of deterministic CBF constraints via the scenario based methodology. Capitalizing on results on scenario based MPC, we provide distribution-free, \emph{a priori} guarantees on the system's closed loop expected safety violation frequency. We demonstrate our results through a case study on unmanned aerial vehicle collision-free position swapping, and provide a numerical comparison with recent stochastic CBF formulations.

\end{abstract}

\section{Introduction}

Ensuring safe interactions between autonomous systems and humans in shared environments is increasingly relevant. Safety formulation in real-time can often be defined through forward set invariance, where control actions keep the system within a safe region over time. An emerging method using this concept is Control Barrier Functions \cite{ames2019control}, which has demonstrated interesting results \cite{ames2014control}, \cite{agrawal2017discrete}. CBFs provide a principled approach to guarantee safety, while being suitable across various applications, including deterministic robotic systems settings \cite{ames2019control}, guaranteeing safety of learning methods \cite{zhao2023stable}, and providing safety assurances under stochastic conditions \cite{clark2021control}, which is relevant in uncertain environments.

A typical safety formulation uses Quadratic Programming (QP), aiming to determine the least restrictive control actions. In particular, the norm of the difference between the control effort and some desired nominal controller \cite{ames2014control}, constrained by CBF inequalities is solved at each time instance. CBF usage has also been investigated in uncertain environments under the robust \cite{xu2015robustness} and stochastic lenses \cite{clark2021control}.

While CBFs are computationally efficient for safety analysis, they can be short-sighted or overly conservative in control invariance calculations. To address this, predictive safety filters have emerged as a promising alternative, often used as an ``add-on" to existing control strategies \cite{wabersich2021predictive}. These approaches leverage model predictive control (MPC) to produce smoother trajectories, reduce safety filter interventions, and lower control effort \cite{tearle2021predictive}, \cite{breeden2022predictive}, \cite{wabersich2021probabilistic}. However, these benefits come with increased computational demands, especially as the prediction horizon increases.

In uncertain or disturbed environments, the advantages of MPC in enforcing safety and managing uncertainty become more evident. To leverage these benefits, we propose a safety formulation that combines the strengths of MPC and CBFs. Our approach co-designs high performance and safe controllers, by solving a finite horizon optimization problem at each time step, like MPC, but only enforcing probabilistic safety constraints via CBFs at the first step. 

We address probabilistic CBF constraints using a scenario-based approach, requiring satisfaction of a finite number of CBF constraints for different scenarios. This data-driven method avoids assumptions about the underlying uncertainty's distribution or set geometry. Our distribution-free formulation builds on scenario-based MPC results \cite{schildbach2014scenario} and provides \emph{a priori} guarantees on the expected average number of safety violations in the closed-loop MPC control sequence.
We demonstrate our method's efficacy on a multi-UAV position swapping and collision avoidance case study and also compare its performance with a state-of-the-art method that uses super martingale principles \cite{cosner2023robust} to bound the probability of exiting the safe set.

The remainder of this paper is structured as follows: Section \Romannum{2} covers results on stochastic and scenario-based MPC. In Section \Romannum{3}, we discuss the notion of probabilistic safety and its guarantees. Section \Romannum{4} illustrates our findings on a multi-UAV position swapping problem, and compares it with a state-of-the-art stochastic CBF method \cite{cosner2023robust}. Section \Romannum{5} provides concluding remarks and future directions.

\section{Related results on MPC}
We consider the design of a real-time safe controller for discrete time linear systems. At each time-step a finite horizon stochastic model predictive control (SMPC) problem will be formulated. The first component of the associated optimal input sequence will be applied to the system and the horizon will be rolled as typically performed in MPC. The first input component of each sequence will be subject to probabilistic safety constraints encoded by chance constraints. As such, the closed-loop input sequence (the concatenation of the first input components of each finite horizon problem) will be (probabilistically) safe by construction. To deal with the chance constraints present in the SMPC setting we follow a scenario based approach. To formalize this procedure we first revisit some results on stochastic and scenario based MPC.

\subsection{Stochastic Model Predictive Control}

Consider a discrete time linear system of the form
\begin{equation}
\label{error_dyn}
x_{t+1}=A(d_t) x_{t}+B_u(d_t) u_{t}+B_{d}(d_t) d_{t},
\end{equation}
where at each time instance $t \in \mathbb{N}$, $x_t \in \mathbb{R}^n$ denotes the state and $u_t \in \mathbb{R}^p$ the control input. The system is subject to a disturbance $d_t \in \mathbb{R}^d$ that affects the system dynamics in an affine manner. Matrices $A(d_t), B_u(d_t)$ and $B_d(d_t)$ are of appropriate dimension.
For each $t$, we assume that $d_t \in D$, where $D$ is endowed with a $\sigma$-algebra $\mathcal{D}$. Let $\mathbb{P}$ denote a probability measure over $\mathcal{D}$ according to which $d_t$ is distributed, and let $\mathbb{E}$ denote the associated expected value operator. 

We consider the following SMPC formulation \cite{cannon2009probabilistic}, \cite{mesbah2016stochastic}, where at each time instance $t$, the finite horizon optimization problem is defined with horizon $N$:
\begin{equation}
 \begin{aligned}
 \label{SMPC}
\min_{u_{0|t}, \dots, u_{N-1|t}} & \sum^{N-1}_{k=0} \mathbb{E}[l(x_{k|t},u_{k|t})]\\
 \textrm{subject to } \  & x_{k+1|t} = A(d_{k|t}) x_{k|t} + B_{u}(d_{k|t})u_{k|t} + B_{d} d_{k|t},\\
 & x_{0|t} = x_t,\\
   &u_{k|t} \in \mathbb{U}, ~\forall k \:=0, \dots, N-1, \\
 & \mathbb{P}[x_{k+1|t} \notin \mathbb{X}, \: ~\forall k \:=0, \dots, N-1] \leq \epsilon.
 \end{aligned}
\end{equation}
Note that $x_{k|t}$ and $u_{k|t}$ denote the state and input, $k$ steps ahead of the current time step $t$ respectively. Our objective is to minimize the expected value of the running cost $l(x_{k|t},u_{k|t})$ over the prediction horizon $k=0, \dots, N-1$. 

The first constraint represents the system dynamics over the horizon. These linear equality constraints can be recursively substituted to eliminate state variables, leaving the chance constraint dependent only on inputs. For clarity, we avoid this substitution.
The second constraint in \eqref{SMPC} is the ``initialization constraint," introducing system feedback at each sampling time $t$. The third constraint imposes bounded inputs, with $\mathbb{U} \subset \mathbb{R}^{p}$. The fourth constraint in \eqref{SMPC} is the chance constraint, limiting the probability of the system state leaving the set $\mathbb{X} \subset \mathbb{R}^{n}$ during the prediction horizon, with a bound $\epsilon$. As a tuning parameter, $\epsilon$ balances robustness and performance: a higher $\epsilon$ reduces the likelihood of staying in $\mathbb{X}$ but improves performance (lower cost).

\subsection{Scenario based Model Predictive Control}

Even if all functions and constraint sets involved are convex, solving \eqref{SMPC} can still be challenging, as chance constraints render the problem in general non-convex. To address this, without assumptions on disturbance distribution or geometry, we adopt a data-driven approach, assuming $D$ and $\mathbb{P}$ are fixed but unknown, with only scenarios (e.g., historical data) $d_t$ available. In the light of this, consider the scenario based MPC problem analyzed in \cite{schildbach2014scenario}: 

\begin{equation}
\begin{aligned}
 \label{SCMPC}
\min_{u_{0|t}, \dots u_{N-1|t}} & \sum^{m}_{i=1} \sum^{N-1}_{k=0} J^i_k(x^{i}_{k|t},u_{k|t}) + J^i_N(x^{i}_{N|t})\\
\textrm{subject to } \ &  x^i_{k+1|t} = A(d^i_{k|t}) x^i_{k|t} + B_u(d^i_{k|t}) u_{k|t} + B_{d}(d^i_{k|t}) d^i_{k|t}, \\
& x_{0|t} = x_t,\\
& x^i_{k+1|t} \in \mathbb{X},\\
& u_{k|t} \in \mathbb{U},\\
& \forall k = 0, \dots, N-1, \quad \forall i = 1, \dots, m. \\
\end{aligned}    
\end{equation}

For each $t, k$, we assume that $d^i_{k|t}$, $i=1,\ldots,m$, are scenarios/samples of $d_{k|t}$ and $J^i_k(x^{i}_{k|t},u_{k|t})$ are cost functions evaluated at $x^{i}_{k|t},u_{k|t}$ where $x^{i}_{k|t}$ is driven by scenario $i$. In this regime we are seeking a sequence of inputs that is consistent with the dynamics and satisfies input and state constraints for all scenarios $i=1,\ldots,m$. However, for a given input, for each scenario a different state trajectory is generated. To reflect this, we introduce superscript $i$ in $x^{i}_{k|t}$ and $d^i_{k|t}$. 

In the sequel, we leverage the scenario-based reformulation of SMPC to generate safe-by-design controllers. Instead of explicitly enforcing safety constraints within $\mathbb{X}$, we adopt an approach using control barrier functions \cite{ames2019control}, \cite{wang2023assessing}. This scenario-based SMPC provides a tractable, though approximate, solution of SMPC, accompanied by probabilistic guarantees on state constraint satisfaction, ensuring safe closed-loop performance.

\section{Problem statement and main results}

\subsection{Scenario based safety}
\subsubsection{Safety constraints}
We consider state constraints encoding safety, as this is represented via control barrier functions. To this end, consider a set $S$ that is defined by the zero super-level set of a function \cite{agrawal2017discrete} denoted as $h:\mathcal{P} \subset \mathbb{R}^{n} \to \mathbb{R}$: 
\begin{equation*}
    S = \{x_t \in \mathcal{P} \subset \mathbb{R}^{n} :h(x_t) \geq 0\} .
\end{equation*}
Ensuring that system trajectories remain within $S$ guarantees safety, which is equivalent to enforcing $S$'s invariance along \eqref{error_dyn} under a given disturbance. Safety conditions can be explicitly derived using discrete-time barrier functions from \cite{agrawal2017discrete}. We assume that $\mathbb{X}$ is defined by the following constraints:
\begin{flalign}
\label{cbf_ineq}
& (i)~ h(x_{0|t}) \geq 0, \nonumber \\
& (ii)~ \exists \: u_{k|t} \text{ such that } \forall t \in \mathbb{N} \cup \{0\}, \\
&\hspace{0.6cm} h(x^i_{k+1|t})-h(x^i_{k|t}) \geq -\gamma h(x^i_{k|t}),~ \forall i = 1, \dots, m \nonumber. &
\end{flalign}
In this context, we assume that $\gamma$ lies within the interval (0, 1), though a more general approach \cite{zeng2021safety} could use a class $\kappa$ functional satisfying $0<\gamma(h(x^i_{k|t})) \leq h(x^i_{k|t})$. Notice that item $(i)$ in \eqref{cbf_ineq} omits the superscript $i$ because, at time $t$ and $k=0$, $h(x_{0|t})$ is unaffected by $d^i_{0|t}$ due to \eqref{error_dyn}. To satisfy item $(ii)$ of \eqref{cbf_ineq}, there must exist at least one $u_{k|t}$ such that the inequality holds for all scenarios $i = 1, \dots, m$ at each look-ahead step $k$. These conditions restrict the solution space, reflecting the ``price of robustness" in a distribution-free setting. When both conditions in \eqref{cbf_ineq} are met, $h: \mathcal{P} \rightarrow \mathbb{R}$ serves as a discrete-time exponential control barrier function for the scenarios.

\subsubsection{Cost function}
For each scenario $i=1,\ldots,m$, the corresponding objective function term in \eqref{SCMPC} is computed as
\begin{align}
\label{cost_fcn}
& J^i_k(x^i_{k|t},u_{k|t}) =  (x^i_{k|t})^T Q x^i_{k|t} + (u_{k|t})^T R u_{k|t}, \nonumber \\
& J^i_k(x^i_{N|t}) =  \eta (x^i_{N|t})^T Q_N x^i_{N|t} 
\end{align}
Matrices $Q \succeq 0$ and $R \succ 0$ penalize the state and control input along the prediction horizon, while $Q_N \succeq 0$ encodes a terminal state penalty, potentially scaled by a high weight $\eta \geq 0$, alleviating the need for a terminal set \cite{limon2006stability}. In multi-agent systems, an additional term may be included to address a team goal, such as a metric for the fleet's ability to follow a shared target \cite{do2023game}. 

\subsection{Closed loop safe set violation guarantees}

Following the developments in \cite{schildbach2014scenario}, we impose the following assumptions on \eqref{SCMPC}.

\begin{assumption} \label{scenaspt}
\begin{enumerate}
    \item \emph{Uncertainty:} Consider the product probability space ($\Delta^{m}$, $\mathbb{P}^{m}$). Assume that for each $t$, the scenarios $[d^i_{0|t},\ldots,d^i_{N-1|t}]$, $i=1,\ldots,m$ are independent and identically distributed elements of the product probability space. 
    \item \emph{Problem structure:} The objective function and state constraint set in \eqref{SCMPC} are convex. The input constraint set is convex and bounded, while full state measurement is assumed.
    \item \emph{Feasibility:} At each instance $t$, it is assumed that problem \eqref{SCMPC} admits almost surely (with respect to the choice of the scenarios) a feasible solution.
\end{enumerate}
\end{assumption}

The first part of this assumption requires scenario independence, but allows temporal correlation within each scenario $i$, e.g., the elements of $[d^i_{0|t},\ldots,d^i_{N-1|t}]$ can be correlated for fixed $i$. The second part is mild, not imposing any specific structure on the system's dependency on uncertainty or disturbance distribution. The third part ensures recursive feasibility for almost all scenarios. Although challenging to satisfy, this requirement can be relaxed while maintaining probabilistic safety guarantees if the problem is feasible \cite{calafiore2006scenario}, \cite{margellos2015connection}.

Let $\omega^i_t = \{d^i_{k|t}\}_{k=0}^{N-1}$ and $\omega_t = \{\omega^i_t\}_{i=1}^m$ be the collection of all finite-horizon samples. 
Given a current state $x_t(\omega_t)$ (notice the dependence on all scenarios up to that time), denote the next-step safety violation probability as $\mathbb{P}[x_{t+1} \notin \mathbb{X}|x_t(\omega_t)]$. Define $\rho$ as the support rank of the constraint $\mathbb{X}$ \cite{schildbach2014scenario}, given by $\rho = p - \text{dim} \mathcal{L}$, where $\text{dim} \mathcal{L}$ is the dimension of the largest unconstrained subspace within the decision variable space of dimension $p$. If the safety constraints are linear (as in our examples), they can be represented as $A_t u_{0|t} \leq b(d_t)$, where $A_t$ is time-dependent. If $A_t \in \mathbb{R}^{z \times m}$ has a rank $r$ for all $d_t$ and all $t$, then $\text{dim} \mathcal{L} = p - r$. In this case $\rho \leq p$ as $0 \leq \text{dim} \mathcal{L} \leq p$. Following \cite[Lemma 13]{schildbach2014scenario}, we can bound this probability with the following lemma.

\begin{lemma}
\label{nxt_stp_prob}
Consider Assumption \ref{scenaspt}. Fix $\epsilon \in (0,1)$, and a confidence level $\beta \in (0,1)$.
Select the number of scenarios $m$ such that 
\[
\min \Big \{1,\sum^{\rho-1}_{j=0} {m \choose j} \epsilon^j (1-\epsilon)^{m-j} \Big \} \leq \beta,
\]
We then have that for each $t$, 
\begin{align}
\label{next_exit}
    \mathbb{P}^{mN}&[\omega_t:~ \mathbb{P}[x_{t+1} \notin \mathbb{X}|x_t(\omega_t)] \leq \epsilon] \geq 1- \beta.
\end{align}
\end{lemma}

In words, for a properly chosen number of scenarios $m$, we guarantee that with confidence at least equal to $1-\beta$, the probability of being unsafe at the next step is at most equal to a prespecified level $\epsilon$.

Lemma \ref{nxt_stp_prob} provides a bound on the next-step probability of being safe. While interesting on its own, in closed-loop operation, we seek guarantees on the safety violations of the closed-loop input sequence. Following the rationale of \cite[Theorem 16]{schildbach2014scenario}, we consider the expected value of the average frequency of these violations. Let $T \in \mathbb{N}$ be the number of MPC horizon rolls, then the average number of safety violations for the closed-loop input sequence is
\[
\frac{1}{T}\sum^{T-1}_{t=0} \mathbf{1}_{\mathbb{X}^c}(x_{t+1}), 
\]
where $\mathbf{1}_{\mathbb{X}^c}$ is an indicator function, which takes the value $1$ if $x_{t+1}$ belongs to the complement of the safe set, namely, $\mathbb{X}^c$, and $0$ otherwise. We then have the following result.

\begin{theorem}
\label{freqtheo}
    Consider Assumption \ref{scenaspt} and fix $\epsilon \in (0,1)$. Select the number of scenarios $m \geq \frac{\rho}{\epsilon} - 1$, where $\rho$ denotes the support rank of the safety constraints. 
    We then have that
    \begin{equation}
    \label{expectedCL}
      \mathbb{E}^{T(mN+1)} \left[\frac{1}{T}\sum^{T-1}_{t=0} \mathbf{1}_{\mathbb{X}^c}(x_{t+1}) \right] \leq \epsilon.
    \end{equation}
\end{theorem}
We believe Theorem \ref{freqtheo}, as stated and proved in \cite{schildbach2014scenario}, is for the first time being used to probabilistically characterize safety violations for closed-loop input sequences generated by MPC and CBF, despite not requiring knowledge of the uncertainty distribution. Both Lemma \ref{nxt_stp_prob} and Theorem \ref{freqtheo} could be extended to allow a fixed number of samples to be removed \emph{a posteriori}, trading feasibility for improved optimality and cost. We do not explore this here but refer to \cite{romao2020tight} and \cite{campi2011sampling} for such approaches. 

\begin{algorithm}
\begin{algorithmic}[1]
\caption{Probably Safe Scenario MPC - (PSS-MPC)}\label{alg:PSSMPC}
\FOR{$t=0,1, \dots$}
    \STATE Measure state $x_t$.
    \STATE Calculate safety constraints as in \eqref{cbf_ineq}. 
    \STATE Fix the values of $\epsilon$ and calculate the support rank $\rho$.
    \STATE Use Theorem \ref{freqtheo} to obtain the number of scenarios $m$.
    \STATE Generate (or collect) $m$ scenarios.
\STATE Solve problem \eqref{SCMPC}, using \eqref{cbf_ineq} and \eqref{cost_fcn}.
\STATE Apply $u_t = u_{0|t}$ in \eqref{error_dyn}.
\ENDFOR
\end{algorithmic}
\end{algorithm}
The procedure for combining CBFs and scenario-based MPC is outlined in Algorithm 1. Step 3 updates safety constraints using the latest $x_t$ from Step 2, assuming the support rank of $h(x_t)$ is unaffected by scenario $i$, as in \cite{schildbach2014scenario}. Steps 4-6 generate the minimum number of scenarios for the chosen $\epsilon$ and $\rho$. If Step 3’s safety constraints are fixed and $\rho$ remains constant, Algorithm 1 runs faster by moving Steps 3-6 out of the loop. Generally, fixing any two of $\epsilon$, $\rho$, or $m$ allows finding the third. Even with limited scenarios $m$, Theorem \ref{freqtheo} can estimate a lower bound $\epsilon$ given $m$ and $\rho$. 

\section{Numerical examples}

\subsection{Position swapping and collision avoidance of UAVs}
\begin{figure}[t]
\centering
\includegraphics[width=0.46\textwidth]{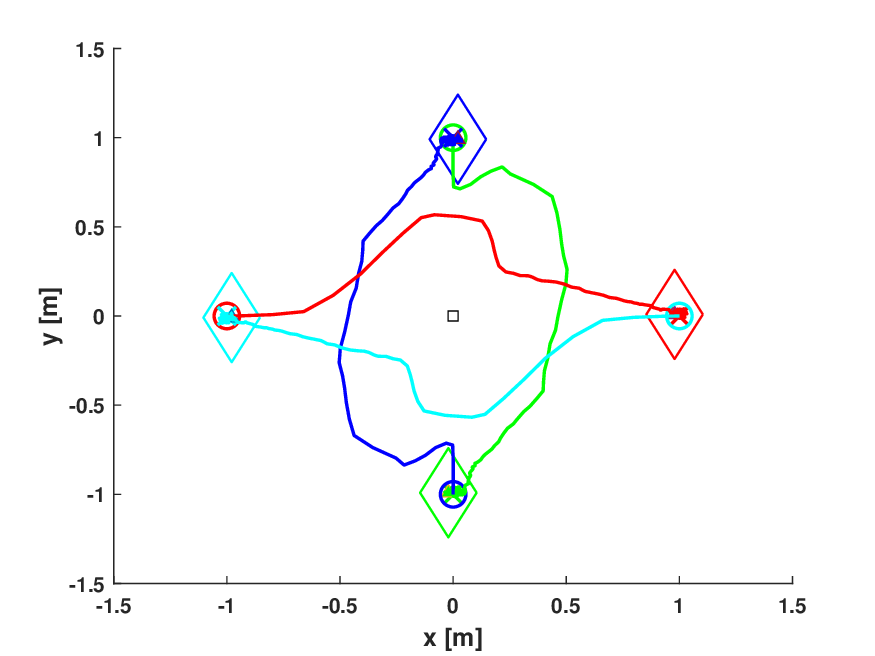}\\
\caption{\label{UAVsnapshot} 
UAV position swap maneuver: polygons show unsafe regions, triangles are UAVs, circles mark initial positions, crosses the targets, and lines depict the closed-loop UAV trajectories.
}
\end{figure}

In \cite{do2023game}, a distributed algorithm for multi-agent UAVs was proposed, resulting in a collision-free control law that enabled position swapping. While we are not focusing on distributed implementation, the proposed scenario program is compatible with such schemes. Instead, we account for disturbances affecting UAV dynamics. Consider four UAVs executing planar trajectories (Figure \ref{UAVsnapshot}). For each $j=1,\ldots,4$, $p_{z_j} = [p_{x_j}, p_{y_j}]$, where $z=\{0,d\}$ denote their initial and desired position. Components $p_{x_j}$ and $p_{y_j}$ represent their horizontal and vertical coordinates. Initial positions are $p_{0_1} = [0, 1]^T$, $p_{0_2} = [0, -1]^T$, $p_{0_3} = [1, 0]^T$, and $p_{0_4} = [-1, 0]^T$. Target positions are set as $p_{d_1} = p_{0_2}$, $p_{d_2} = p_{0_1}$, $p_{d_3} = p_{0_4}$, and $p_{d_4} = p_{0_3}$. 

\emph{Dynamics:} All UAVs are represented as discrete-time double integrators; for each $j=1,\ldots,4$, each UAV's state is given by $x_j=[p_{x_j},p_{y_j},v_{x_j},v_{y_j}]^T$, where $v_{x_j}$ and $v_{y_j}$ denote the horizontal and vertical velocity components. 
These dynamics are affected by an additive disturbance as in \eqref{error_dyn}, 
$B_{d}^j = 5 \times 10^{-3} \times (-1)^{j+1} [I_{2\times2};0_{2\times2}]$, where $I_{2\times2}$ and $0_{2\times2}$ denote the $2 \times 2$ identity and zero matrices. The superscript $j$ denotes the matrices/vectors for UAV $j$. We assume that all disturbance scenarios are independently drawn from a normal distribution $\mathcal{N}(0,1)$, with samples also independent across the horizon and from the same disturbance type. 

\emph{Objective function:} All agents $j=1,\ldots,4$ share the same matrices in the objective function: $Q^j = 5 \times I_{4 \times 4}$, $R^j = 2 \times I_{2 \times 2}$, and $Q^j_{N}$ is set by the discrete-time algebraic Riccati equation. We also set $\eta = 0.1$. The overall system matrices $Q$, $R$, and $Q_N$ in \eqref{cost_fcn} are formed by diagonally concatenating $Q^j$, $R^j$, and $Q^j_{N}$, respectively. 

\emph{Constraints:} Safety in this setting is encoded by avoiding all pairwise collisions, enforced by the two conditions in \eqref{cbf_ineq}. A discrete time CBF to achieve this was detailed in \cite{do2023game}; with a slight abuse of notation, the CBF $h_{i,j}$ between agents $i$ and $j$, is $h_{i,j} = \frac{\lvert p_{x_i} - p_{x_j}\rvert}{r_1} + \frac{\lvert p_{y_i} - p_{y_j}\rvert}{r_2} - 1$, where $r_1$ and $r_2$ are collision radii set to $0.25 m$ and $0.5m$, and we set $\gamma = 0.2$. This CBF was linearized around the states $x_i$, $x_j$ at time $t$, giving rise to constraint $A_t u_{0|t} \leq b(d_t)$.

The UAV acceleration input is constrained by $-a_{\max} \times 1_{8 \times 1} \leq u_{k|t} \leq a_{\max} \times 1_{8 \times 1}$, where $1_{8 \times 1}$ is a ${8 \times 1}$ column vector of ones, $u^j_{k|t} \in \mathbb{R}^{2 \times 1}$ is the $k$ steps ahead input calculated for UAV $j$ at time $t$, $u_{k|t} = [(u^1_{k|t})^T, \dots, (u^4_{k|t})^T]^T$ and $a_{max} = 4 m/s^2$.

\emph{Scenario based MPC and probabilistic guarantees:} To solve the proposed scenario based MPC we used a prediction horizon $N = 3$, for better performance while maintaining safety as explained in \cite{do2023game}. The matrix $A_t$ has rank $6$ (full row rank) as long as the system is safe. As such, its support rank is $\rho = 6$. Theorem \ref{freqtheo} implies that to meet the level $\epsilon = 0.05$, we need at least $m=119$ scenarios.

\begin{figure}[t]
\centering
\includegraphics[width=0.46\textwidth]{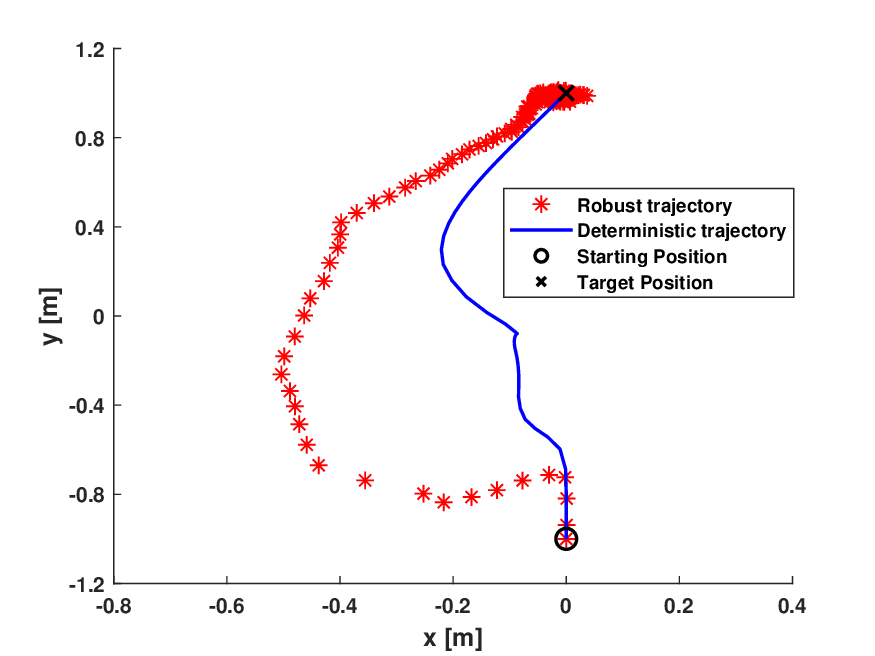}\\
\caption{\label{UAVtrajcomp} Deterministic (blue) and robust (solution of the scenario based MPC) trajectories (red) for one UAV of Fig.\ref{UAVsnapshot}. 
}
\end{figure}

The resulting trajectories of the scenario based MPC problem are shown in Figure \ref{UAVsnapshot}. These trajectories achieve position swapping, and UAVs avoid collision for unseen (albeit of the same nature) disturbance, by being robust with respect to the considered scenarios. This is seen as the minimum CBF value for all pairwise UAVs during their fight is $h_{1,2} = 0.0054$. This means UAVs 1 and 2 despite being the closest to unsafety, are still safe as $h_{1,2} \geq 0$. At the same time, by Theorem \ref{freqtheo}, the expected value of the average number of safety violations is at most $0.05$. For one of the considered UAVs, Figure \ref{UAVtrajcomp} contrasts the deterministic trajectory with the designed (probabilistically) robust one. The associated trajectories for the other UAVs are qualitatively similar.

\begin{figure}[t]
\centering
\includegraphics[width=0.46\textwidth]{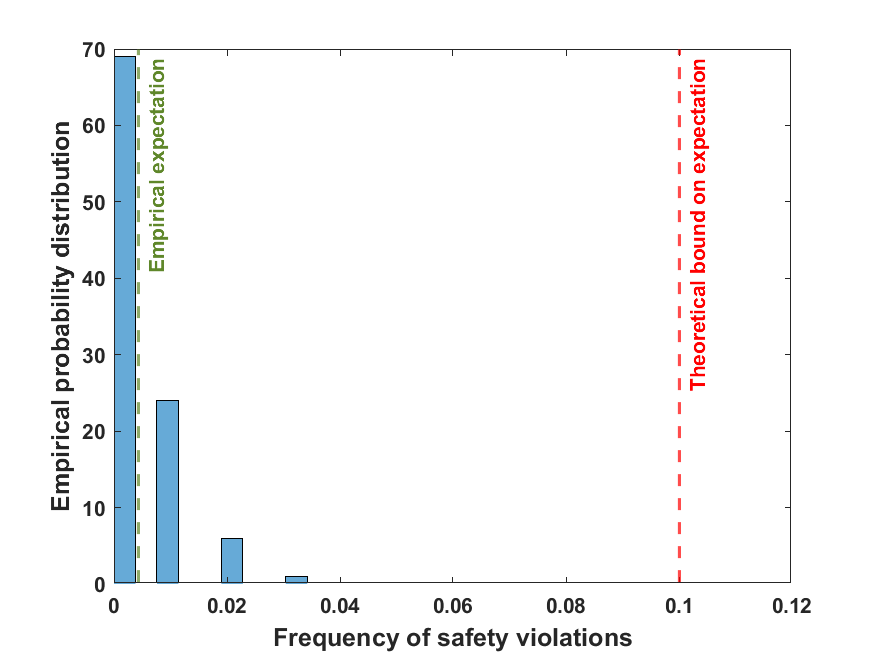}\\
\caption{\label{expected_freq} Empirical distribution (calculated based on $100$ independent runs) of average safety violations for $T=90$ time steps.}
\end{figure}

We empirically validate Theorem \ref{freqtheo} by setting $\gamma = 0.9$ and $\epsilon = 0.1$, which requires generating $m = 59$ scenarios according to the theorem. Over $T = 90$ time steps, we calculate the empirical expectation $\widehat{\mathbb{E}}$ for the left-hand side of \eqref{expectedCL}. At each time step, we generate $m = 59$ scenarios, solve the scenario program, and apply the resulting input $u^j_{0|t}$ to each UAV ($j = 1,\ldots,4$). We compute the empirical frequency of safety violations across these time steps, considering any number of collisions at a given time as one unsafe instance to avoid double counting. This process is repeated for $M = 100$ independent runs, recording the empirical frequency $f_\ell$ for each run. Finally, we construct the empirical expectation $\widehat{\mathbb{E}}$  
\[
\widehat{\mathbb{E}} = \frac{1}{M}\sum^{M}_{\ell=1} f_\ell.
\]
Figure \ref{expected_freq} shows the empirical distribution of average safety violations over $T = 90$ time steps. The dashed red line represents the theoretical bound $\epsilon$ (from Theorem \ref{freqtheo}), while the dashed green line shows the empirical expectation $\widehat{\mathbb{E}}$. Using a sampling and discarding mechanism \cite{romao2020tight}, \cite{campi2011sampling} could reduce conservatism and align the empirical expectation with the theoretical one. We chose $\gamma = 0.9$ to test the theorem's bound in \eqref{expectedCL} with less conservatism (on the barrier function side), as lower $\gamma$ values make the system more cautious, reducing safety violations and the empirical expectation.


\subsection{Comparative study}
We compare our probabilistic safety guarantees with \cite{cosner2023robust} using the same numerical example as in \cite{cosner2023robust}. Consider a one-dimensional linear system: $x_{t+1} = x_t + 2 + u_t + \sigma d_t$, where $d_t \mathtt{\sim} \mathcal{N}(0,1)$, and define the safe set as $S_{\nu}=\{x|h(x) \geq -\nu\}$, with $h(x) = 10-x^2$ and $\nu = 0$. 

The methodology in \cite{cosner2023robust} uses a CBF tightening procedure and martingale arguments to ensure the controller's probability of exiting the safe set at some point in time. While our results differ, offering bounds on the expected average constraint violations of the closed-loop system, our main theorem's bound relies on Lemma \ref{nxt_stp_prob}, which estimates the one-step probability of exiting the safe set. This shared focus on the one-step exit probability allows us to compare these approaches. For the mentioned example, \cite{cosner2023robust} considers the following CBF problem
\begin{align}
\label{orig}
    u_t = \: \text{argmin} &\lVert u \rVert^2 \nonumber\\
    \textrm{subject to } \ &  \mathbb{E}[h(x_{t+1})] \geq (1-\sigma^2) h(x_t).
\end{align} 
As the expectation operator cannot be parsed inside $h$ (apart from specific cases), the following tightening version is considered
\begin{align}
\label{JED}
    u_t = \: \text{argmin} & \lVert u \rVert^2 \nonumber\\
    \textrm{s.t.} \ & h(\mathbb{E}[x_{t+1}]) -\sigma^2 \geq (1-\sigma^2) h(x_t).
\end{align}
Recall that $\sigma$ is the coefficient of $d_t$ in the dynamics and effectively plays the role of standard deviation of the term $\sigma d_t$.
It is shown that the solution of the tightened problem is also a feasible solution to \eqref{orig}. The one-step exit probability bound proposed in \cite{cosner2023robust}, when adapted to this problem is: 
\begin{equation}
\label{boundJed}
    P_u \leq 1-\frac{h(x_0)}{M_{cbf}} (1-\sigma^2),
\end{equation}
where $M_{cbf}$ represents the maximum value of $h(x)$ and $h(x_0)$ is its initial value. It is worth noting that the bound becomes increasingly close to 1 as the system approaches the safe set boundary. 

In our approach, Lemma \ref{nxt_stp_prob} suggests that the one-step exit probability is given by $\epsilon$. We can thus fix any $\epsilon$ as long as the number of samples $m$ and the confidence level $\beta$ are chose so that $\min \Big \{1,\sum^{\rho-1}_{j=0} {m \choose j} \epsilon^j (1-\epsilon)^{m-j} \Big \} \leq \beta$.

\begin{figure}[t]
\centering
\includegraphics[width=0.46\textwidth]{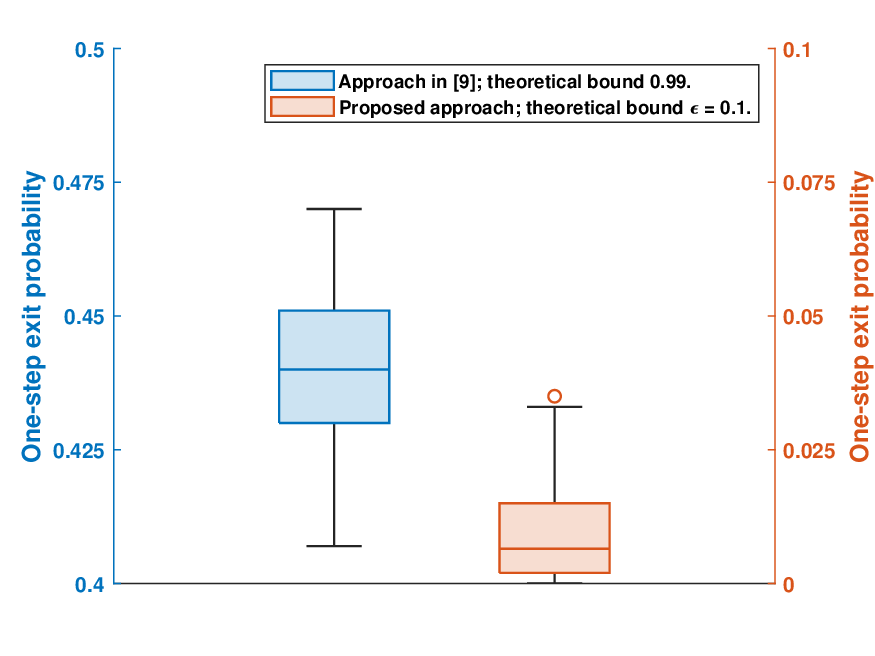}\\
\caption{\label{boxplotScomp} Comparison of one-step exit probability:  \cite{cosner2023robust} and ours (Lemma \ref{nxt_stp_prob}). The left boxplot, linked to the left verical axis represents the approach in  \cite{cosner2023robust}. The Right boxplot, linked to the right vertical axis corresponds to our proposed approach.}
\end{figure}

We set $\sigma=0.9$, $x_0 = 3.1$, and $\epsilon = 0.1$, requiring $m=88$ samples for a confidence level $\beta=10^{-4}$. For these figures, the theoretical upper-bound on the one-step exit probability for \cite{cosner2023robust} is $P_u \leq 0.99$ (follows from \eqref{boundJed}) while for ours is $\epsilon = 0.1$.
Figure \ref{boxplotScomp} provides a statistical analysis for this comparison. We have simulated the system under consideration using the controller emanating from \eqref{JED} and using the controller generated by our scheme, setting $T=1$ time step. For each controller, we counted the empirical frequency of unsafe occurrences out of $1000$ of these time steps. This process was repeated for $M=100$ independent runs. The distribution of empirical frequencies is illustrated by two boxplots in Figure \ref{boxplotScomp}. The probabilistic bounds observed for the methodology of \cite{cosner2023robust} are concentrated to higher values, thus making our approach's guarantees sharper.

While empirical calculations do adhere to the theoretical bound, it has come to our attention that controller \eqref{JED} sometimes encounters difficulties in leveraging its unbounded input capacity to ensure system safety. In contrast, the scenario-based controller maintains the system within the safe set for over $90\%$ of the time with arbitrary confidence, irrespective of its proximity to the boundary. This suggests it presents an efficient utilization of control capacity, despite the smaller size of the feasible set arising from the numerous constraints generated by the scenarios and its increased computation time.

\addtolength{\textheight}{-3cm}   


\section{Concluding remarks and future work}

In this study, we introduce a sample-based approach for safe model predictive control, bridging the gap between stochastic QP-CBF and safety filters. Our method avoids the pointwise safety guarantees of myopic controllers without adding excessive constraints. It provides upper bounds on the expected frequency of chance constraint violations for systems in closed-loop by integrating a scenario-based MPC framework with control barrier functions. We evaluate this approach through two numerical studies: a UAV swapping task under external disturbances and a comparison with a state-of-the-art method offering future step exit probability bounds. These studies focus on next-step safe set exits. Promising future research includes extending the ``CBF horizon" for non-convex safe sets, ensuring recursive feasibility under mission-wide probabilistic constraints as in \cite{wang2021recursive}, investigating probabilistic safety under changing disturbance profiles and how higher order CBFs could impact the method proposed. Additionally, the methods compared in Section IV have interesting similarities prompting further exploration.



\bibliographystyle{IEEEtran}

\bibliography{refs}

\end{document}